\documentclass[aps,prd,amsfonts,preprint, eqsecnum,nofootinbib]
{revtex4} \voffset  -0.1in %\addtolength{\textheight}{-0.25in}
\usepackage{graphics,epsfig}

\begin{document}
\preprint{WM-05-126}

\count255=\time\divide\count255 by 60 \xdef\hourmin{\number\count255}
  \multiply\count255 by-60\advance\count255 by\time
 \xdef\hourmin{\hourmin:\ifnum\count255<10 0\fi\the\count255}

\title{Diquark Correlations from Nucleon Charge Radii}

\author{Carl E. Carlson}
\email{carlson@physics.wm.edu}
\author{Christopher D. Carone}
\email{carone@physics.wm.edu}
\author{Herry J. Kwee}
\email{hxjohn@wm.edu}

\affiliation{Particle Theory Group, Department of Physics, College of William \& Mary,
Williamsburg, VA 23187-8795}

\author{Richard F. Lebed}
\email{Richard.Lebed@asu.edu}

\affiliation{Department of Physics and Astronomy, Arizona State
University, Tempe, AZ 85287-1504}

%\date{\hourmin, \today}
\date{December, 2005}

\begin{abstract}
We argue that precise measurements of charge and magnetic radii can meaningfully constrain
diquark models of the nucleon.  We construct properly symmetrized, nonrelativistic three-quark
wave functions that interpolate between the limits of a pointlike diquark pair and no diquark
correlation.  We find that good fits to the data can be obtained for a wide range of diquark
sizes, provided that the diquark wave functions are close to those that reduce to a purely scalar
state in the pointlike limit. A modest improvement in the experimental uncertainties will render a
fit to the charge radii a more telling diagnostic for the presence of spatially correlated quark
pairs within the nucleon.
\end{abstract}

%\pacs{14.20.Dh, 13.40.Gp, 12.39.Jh}
%14.20.Dh   Protons and neutrons
%13.40.Gp   Electromagnetic form factors
%12.39.Jh   Nonrelativistic quark model

\maketitle

\section{Introduction} \label{intro}

``Are there correlations among quarks?'' is a simple yet central
question in the study of hadron structure.   One conjecture that
has received attention over many years~\cite{Fredriksson:1981za,Haire:1983ff,Kroll:1991xm},
and renewed advocacy in recent years~\cite{Jaffe:2003sg,Wilczek:2004im}, is
that pairs of quarks form identifiable ``diquark'' substructures,
where diquarks have a smaller size scale than hadrons overall and
have definite spin and isospin quantum numbers. The name
``diquark'' has been used with various meanings.  It sometimes refers simply
to the two quarks remaining after one quark is scattered out of a
baryon~\cite{Sukhatme:1981ym}; of more relevance to the
present investigation, it has been used to refer to a pointlike or
nearly pointlike state with the quantum numbers of two
quarks~\cite{Fredriksson:1981za,Haire:1983ff,Kroll:1991xm}.   We
wish to learn if experimental evidence indicates the presence of diquarks
that are small, but not necessarily pointlike, two-quark correlations. At some
level, the existence of quark-quark correlations has been known for decades, since the
nucleon-$\Delta$(1232) mass splitting is generally attributed to a
spin-spin interaction~\cite{DeRujula:1975ge} that is attractive for
spin-zero isoscalar quark pairs and repulsive for spin-one
isovector pairs.  The attractive interaction shrinks the size
of the favored quark pairs, but is unlikely to force them to be near
the pointlike diquark limit.

In this work we are interested in nucleon form factors, which provide
direct information on hadron structure.  Nucleon form factors have been
studied previously in a diquark-quark model~\cite{Jakob:1993th}, but
the diquark has in these studies been taken as a distinct state,
sometimes including its own intrinsic form factor.  Other recent work on the
consequences of diquarks in hadron phenomenology can be found in Ref.~\cite{other}.
Here we start with a three-quark state and describe the nucleon with wave
functions that contain quark-quark correlated states bound to a
third quark.  We include scalar diquarks, which refer to
isospin-zero states, which must also be spin-zero if bound in a
relative S-state, and vector diquarks, which are the corresponding
spin-one isospin-one possibilities.  We do not overlook the role of Fermi
statistics.  If the diquark state is very small, the
effects of antisymmetrizing the state with respect to interchanges
of one of the quarks in the diquark and the third quark may be
neglected. Many authors have yielded to the temptation of assuming a
simplified hadronic state that is not totally antisymmetric.   However, if the
diquark is not much smaller than the hadron, overall
antisymmetrization is important.  All of our states
are completely antisymmetric among all three quarks. We focus in
this note on the charge and magnetic radii of the proton and neutron.  These
are low-momentum transfer quantities and allow us to use
nonrelativistic models.  Further, we limit ourselves to
three-quark Fock components of the nucleons, without relative orbital
angular momentum.  Our wave functions
are written in Jacobi coordinates, which naturally allow a
quark-quark correlated state of one length scale bound to the third
quark with a wave function of a different and presumably larger
length scale.  Scalar and vector diquarks in an arbitrary mixture are allowed.
The formal details of our calculation are given in Sec.~\ref{formal}, and the
results of our fits, including the allowed scalar-vector diquark mixture
and the ratio of length scales, are presented in Sec.~\ref{models}.  Section~\ref{concl}
summarizes our conclusions.

\section{Formalism} \label{formal}

The nucleon wave function is completely symmetric under
spin$\times$flavor$\times$space, with color, as usual, carrying the
antisymmetry necessary to guarantee the proper fermionic statistics.
The flavor and spin wave functions are of the standard textbook
forms~\cite{HM}, for example,
\begin{equation}
\begin{array}{lcl}
p_{\rm MS} =  \frac{1}{\sqrt{6}} [2 uud -(ud + du) u] \ , & & p_{\rm MA} =  \frac{1}{\sqrt{2}}
(ud - du)u  \ , \\ \chi_{\rm MS} (\uparrow) =  \frac{1}{\sqrt{6}} [2\uparrow\uparrow\downarrow
-(\uparrow\downarrow + \downarrow\uparrow) \uparrow] \ , & \mbox{\hspace{3em}}&
\chi_{\rm MA} (\uparrow)  = \frac{1}{\sqrt{2}}
( \uparrow \downarrow - \downarrow \uparrow ) \uparrow \ .
\end{array}
\end{equation}
The mixed-symmetric (MS) wave functions are isovector (vector) in flavor (spin)
and are symmetric in the first two quark indices.  Likewise, the mixed-antisymmetric (MA)
wave functions are isoscalar (scalar) and antisymmetric in the first two quark indices.
One may construct a nucleon state that contains a pure scalar (vector) diquark by multiplying
$p_{\rm MA} \chi_{\rm MA}$ ($p_{\rm MS} \chi_{\rm MS}$) with a wave function
$\phi_{\rm MS}$---either in position or momentum space---that is also symmetric under exchange of
the first two quarks.  Such a combined wave function is, however, not completely symmetric under
{\em all\/} quark exchanges because it treats the first two quarks as special; one must
add to it a similar expression that distinguishes quark pair $\{13\}$ and one that
distinguishes quark pair $\{23\}$.

We denote the completely symmetric wave function obtained in this way from
$p_{\rm MA} \chi_{\rm MA}\phi_{\rm MS}$ as $|S\rangle$, and the symmetrized wave function
obtained from $p_{\rm MS} \chi_{\rm MS} \phi_{\rm MS}$ as $|V\rangle$. Significantly, in
general $\langle S |V \rangle \! \neq \! 0$.  Taking $| 1 \rangle$ to be the normalized
form of $| S \rangle$, we define a state $|2\rangle$,  via the Gram-Schmidt process,  that
is proportional to $|V\rangle$ minus its projection along $|S\rangle$, so that
$\langle 1 | 2 \rangle \! = \! 0$ and $\langle 1 | 1 \rangle \! = \!\langle 2 | 2 \rangle  \! = \!1$ .
Finally, the full wave function $| \psi \rangle$ consists of a linear combination of these states:
\begin{equation}
| \psi \rangle \equiv \cos \theta \; | 1 \rangle + \sin \theta \; | 2
\rangle \ .
\end{equation}
An overall sign flip in this wave function is unobservable, and
therefore all observables obtained depend only upon quadratic forms in
$\cos \theta$, $\sin \theta$, or equivalently, upon $\sin 2\theta$ and
$\cos 2\theta$.

We employ a notation that can be applied to any choice of the nonrelativistic
3-quark spatial or momentum space wave function.   We begin by assuming only that the
momentum-space wave function in the center of mass frame may be written as a
function $f$ of the form
\begin{equation} \label{form}
\phi ( a b c ) = f ( p_{-}^{(abc)} ,  p_{+}^{(abc)}) \ ,
\end{equation}
where ($abc$) represents the three single-particle coordinates in any
fixed order, and the momenta in Jacobi coordinates are defined as
\begin{eqnarray}
{\bf p}_{-}^{(abc)} & \equiv &
\frac{1}{\sqrt 2} ( {\bf p}_a - {\bf p}_b ) \ , \nonumber \\
{\bf p}_{+}^{(abc)} & \equiv &
\frac{1}{\sqrt 6} ( {\bf p}_a + {\bf p}_b - 2 {\bf p}_c ) \ .
\end{eqnarray}
Note that $\phi (a b c) \! = \! \phi (b a c)$.

First we define the basic overlap integrals
\begin{eqnarray}
M   & \equiv & \langle \phi( a b c ) \left| \right. \phi ( a b c )
\rangle \ , \nonumber \\
\mu & \equiv & \langle \phi( a b c ) \left| \right. \phi ( a c b )
\rangle \ .
\end{eqnarray}
Of course, if $\phi$ is normalized then $M \! = \! 1$.  Moreover, due
to the symmetry under coordinate exchanges, every other such overlap
integral is equal to $M$, $\mu$, or $\mu^*$.

We are interested in matrix elements
\begin{equation}
\langle \phi( a b c ) \left| \, {\cal O}_d \, \right| \phi ( e f g )
\rangle \ ,
\end{equation}
where in configuration space, ${\cal O}_d \! = \! \exp ( -i \, {\bf Q}
\! \cdot {\bf r}_d )$.  It is straightforward to show, again owing to
symmetry under coordinate exchanges, that only four independent matrix
elements arise:
\begin{eqnarray} \label{ints}
\Delta_1 & \equiv & \langle \phi( a b c ) \left| \, {\cal O}_a \,
\right| \phi ( a b c ) \rangle \ , \nonumber \\
\Delta_2 & \equiv & \langle \phi( a b c ) \left| \, {\cal O}_c \,
\right| \phi ( a b c ) \rangle \ , \nonumber \\
\delta_1 & \equiv & \langle \phi( a b c ) \left| \, {\cal O}_a \,
\right| \phi ( a c b ) \rangle \ , \nonumber \\
\delta_2 & \equiv & \langle \phi( a b c ) \left| \, {\cal O}_b \,
\right| \phi ( a c b ) \rangle \ .
\end{eqnarray}
  Clearly, $\Delta_{1,2} ({\bf Q} \! = \! 0)
\! = \! M$ and $\delta_{1,2} ({\bf Q} \! = \! 0) \! = \! \mu$.  All
other permutations of indices lead to matrix elements that equal
either $\Delta_1$, $\Delta_2$, $\delta_1$, $\delta_1^*$, $\delta_2$,
or $\delta_2^*$.  In the case that $\phi$ is manifestly real, so are
$\mu$ and $\delta_{1,2}$; otherwise, the following expressions remain
true under the replacements $\mu \to {\rm Re} \, \mu$ and
$\delta_{1,2} \to {\rm Re} \, \delta_{1,2}$.

Using Eq.~(\ref{form}), these four integrals may be written as special
cases of the single master integral
\begin{eqnarray}
\lefteqn{{\cal I} ( \gamma , \delta ) = \int \! \! \int
d^3 {\bf p}_{-}^{(abc)} d^3 {\bf p}_{+}^{(abc)}
f^*( | {\bf p}_{-}^{(abc)} | , | {\bf p}_{+}^{(abc)} | )} \nonumber \\
& \times & \! \! f ( | (\cos \gamma) {\bf p}_{-}^{(abc)} \! +
(\sin \gamma) {\bf p}_{+}^{(abc)} \! - \! \sqrt{\scriptstyle \frac 2 3}
(\sin \delta) {\bf Q} | , | (\sin \gamma) {\bf p}_{-}^{(abc)}
\! - (\cos \gamma) {\bf p}_{+}^{(abc)} \! - \!
\sqrt{\scriptstyle \frac 2 3} ( \cos \delta ) {\bf Q} | ) ,
\nonumber \\ \label{eq:masteri}
\end{eqnarray}
where $\Delta_1 \! = \! {\cal I} ( 0, \frac{2\pi}{3} )$, $\Delta_2 \!
= \!  {\cal I} ( 0, 0 )$, $\delta_1 \! = \! {\cal I} (
-\frac{2\pi}{3}, -\frac{2\pi}{3} )$, and $\delta_2 \! = \! {\cal I} (
\frac{2\pi}{3}, 0 )$.

The nucleon electromagnetic form factors
\begin{eqnarray}
G_E (Q^2) & \equiv & \langle \psi | \sum_i e_i {\cal O}_i |
\psi \rangle \ , \nonumber \\
G_M (Q^2) & \equiv & \langle \psi | \sum_i e_i (\sigma_z)_i
{\cal O}_i | \psi \rangle \ ,
\end{eqnarray}
where the $e_i$ are quark charges, can be expressed as
\begin{eqnarray}
G (Q^2) & = & \frac 1 3 \left( \cos \theta - \frac 3 2
\frac{\mu \sin \theta}{\sqrt{(M+2\mu)(M-\mu)}} \right)^2
\frac{{\cal M}_1}{2M+\mu} \nonumber \\ & &
+ \frac 1 3 \left( \cos \theta - \frac 3 2 \frac{\mu \sin \theta}
{\sqrt{(M+2\mu)(M-\mu)}} \right)
\frac{\sin \theta}{\sqrt{(M+2\mu)(M-\mu)}} {\cal M}_2 \nonumber \\
& &  + \frac{(2M+\mu) \sin^2 \theta}{36(M+2\mu)(M-\mu)} {\cal M}_3 \ ,
\end{eqnarray}
where the matrix elements ${\cal M}$ are given by
\begin{equation}
\begin{array}{lcl}
{\cal M}_{1,E}^p  =  2 \Delta_1 + 4 \Delta_2 - \delta_1 + 4 \delta_2 \ , & \mbox{\hspace{2em}} &
{\cal M}_{1,E}^n  =  2( \Delta_1 - \Delta_2 + \delta_1 - \delta_2 )\ , \\
{\cal M}_{2,E}^p  =  3( \delta_1 + 2 \delta_2 ) \ , & &
{\cal M}_{2,E}^n  =  0 \ , \\
{\cal M}_{3,E}^p  =  9( 2 \Delta_1 + \delta_1 ) \ , & &
{\cal M}_{3,E}^n  =  -6( \Delta_1 - \Delta_2 + \delta_1 - \delta_2 )\ , \\
{\cal M}_{1,M}^p  =  4 \Delta_2 + \delta_1 + 4 \delta_2 \ , & &
{\cal M}_{1,M}^n  =  -2( \Delta_2 + \delta_1 + \delta_2 ) \ , \\
{\cal M}_{2,M}^p  =  2 \Delta_1 + \delta_1 + 6 \delta_2 \ , & &
{\cal M}_{2,M}^n  =  -2( \Delta_1 + 2\delta_2 ) \ , \\
{\cal M}_{3,M}^p  =  3( 4 \Delta_1 + 5 \delta_1 ) \ , & &
{\cal M}_{3,M}^n  =  -2( 2\Delta_1 + \Delta_2 + 5\delta_1 + \delta_2) \ .
\end{array}
\end{equation}
From these forms one can show that $G^p_E(0) \! = \! 1$ and $G^n_E(0)
\! = \! 0$, independent of $\theta$.  The charge radii are defined by
\begin{equation}
\left. r^2 \equiv -\frac{6}{G(0)}
\frac{dG(Q^2)}{dQ^2} \right|_{Q^2 = 0} \ ,
\end{equation}
with the denominator $G(0)$ omitted for the electric form factor of
neutral particles.

\section{Models and Results} \label{models}

In the models used in this paper, the scales of the momenta ${\bf
p}_{-}$ and ${\bf p}_{+}$ are set by parameters $\alpha$ and $\beta$
respectively, each having dimensions of length.  All dimensionless
observables, such as the ratios of charge radii, are therefore
functions only of the two dimensionless parameters $\alpha / \! \beta$
and $\cos \theta$.  The small-diquark limit is defined by $\alpha / \!
\beta \! \ll \! 1$ and the no-diquark limit by $\alpha=\beta$.

\subsection{Dipole Model}

We begin by examining a model defined by the normalized momentum-space wave
function
\begin{equation}
\phi ( a b c ) = \frac{8}{\pi^2} ( \alpha \beta )^{3/2}
\left[ 1 + \alpha^2 ({p}_{-}^{(abc)})^2 \right]^{-2}
\left[ 1 + \beta^2  ({p}_{+}^{(abc)})^2 \right]^{-2} \ .
\end{equation}
This form is motivated by its simplicity and by the fact that it
guarantees that the resulting form factors fall off as $1/Q^4$, as predicted
by QCD~\cite{BL}.  As we note in the following subsection, however, this requirement is not
strictly required since we are working within a nonrelativistic framework that may not be applied
reliably to large values of $Q^2$.

%%%%%%%%%%%%%%%%%%%%%%%%%%%%
\begin{figure}[ht]
\begin{center}
\epsfxsize 3.5 in \epsfbox{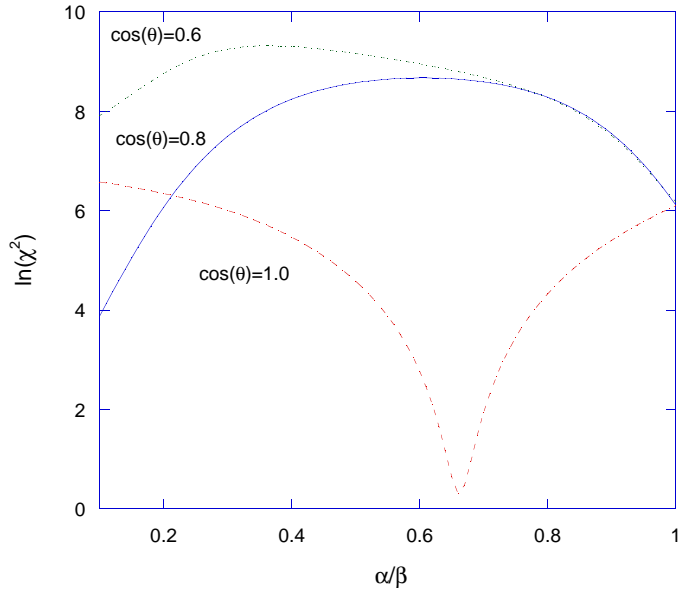} \caption{Plot of
$\ln(\chi^2)$ for fixed mixing angle, assuming dipole wave
functions.\label{fig:dipolefig}}
\end{center}
\end{figure}
%%%%%%%%%%%%%%%%%%%%%%%%%%%%

We have studied the integrals of Eq.~(\ref{ints}) numerically for arbitrary
$\alpha/\beta$ and computed them exactly in the limit $\alpha/ \! \beta \! \to \! 0$.  For
particularly small diquarks, one finds
\begin{eqnarray} \label{abrad}
r_{Ep}^2 & = & ( 2 + \cos 2\theta ) \cdot \frac{\beta^2}{2} \ ,
\nonumber \\
r_{Mp}^2 & = & \frac{5 + 3 \cos 2\theta + \sin
2\theta}{2 + \sin 2\theta} \cdot \frac{\beta^2}{2} \ , \nonumber \\
r_{En}^2 & = & - \cos 2\theta \cdot \frac{\beta^2}{2} \ , \nonumber \\
r_{Mn}^2 & = & \frac{ 3 + \cos 2\theta + \sin 2\theta }
{ 1 + \sin 2\theta } \cdot \frac{\beta^2}{2} \ .
\label{eq:asymptopia}
\end{eqnarray}
In fact, one can shown that any ratio of the squared charge radii above is
independent of the model wave function. Supposing that the $\theta$ dependence
of the charge radii varies weakly in $\alpha/\! \beta$, and noting the basic pattern of the
charge radii data, $r_{E,p}^2 \! \simeq \! r_{M,p}^2 \! \simeq \!
r_{M,n}^2$ and $r_{E,n}^2 \! < \! 0$, one may use Eqs.~(\ref{abrad}) to
predict disfavored regions of $\theta$.  In particular, $r_{E,n}^2 \!
< \! 0$ only for $|\cos \theta | \! > \! 1/\sqrt{2} \! \simeq 0.7$,
while $r_{E,p}^2 / r_{M,p}^2 \! \in [ 0.75, 1.0 ]$ for $-0.4 \! < \!
\cos \theta \! < \! 1.0$ and $r_{M,n}^2 / r_{E,p}^2 \! \in [ 0.93, 1.33
]$ for $0.4 \! < \! \cos \theta \! < \! 1.0$.  One therefore
expects a favored region (noting that $\cos \theta \! = \! \pm 1$ are
the same solution) near $|\cos \theta| \! = \! 1$.  Indeed, this is
borne out by our full numerical simulations.

Numerical evaluation of the integrals in Eq.~(\ref{ints}) was performed
using both Gaussian quadrature and adaptive Monte Carlo integration
techniques.  A $\chi^2$ fit was performed for the three ratios of the squared charge
radii, $r^2_{M,p}/r^2_{E,p}$, $r^2_{E,n}/r^2_{E,p}$, and $r^2_{M,n}/r^2_{E,p}$,
while the overall scale was fixed afterwards using the experimental value of $r^2_{E,p}$. (Nearly
identical results were obtained when the overall scale was included in the $\chi^2$ function.)
There are several determinations of the charge and magnetic radii in the
literature~\cite{HdJ,sick,Kelly,codata}. We have used the radii and uncertainties from
the review of Hyde-Wright and de Jager~\cite{HdJ}, which are summarized in Table~\ref{expdata}.
\begin{table}[ht]
\begin{center}
\caption{Values for the nucleon charge and magnetic radii. \label{expdata}}
\label{radii}
\begin{tabular}{lcc}
\hline \hline
\textbf{Observable} & \hspace{1em} \textbf{value $\pm$ uncertainty} \hspace{1em} & \textbf{Reference}  \\
\hline
$ r_{Ep}$ & $0.895\pm 0.018$~fm & \cite{sick} \\
$ r_{Mp}$ & $0.855\pm 0.035$~fm & \cite{sick} \\
$ r_{En}^2$ & $-0.119\pm 0.003$~ fm$^2$ & \cite{kope} \\
$ r_{Mn}$ & $0.87\pm 0.01$~fm & \cite{kubo} \\
\hline\hline
\end{tabular}
\end{center}
\end{table}

Values of $\alpha/\beta$ between $0.1$ and $1.0$ were sampled and the value
of $\cos\theta$ yielding the lowest $\chi^2$ was determined in each case.
(For smaller $\alpha/\beta$, the asymptotic forms given in Eq.~(\ref{eq:asymptopia})
yield a $\chi^2$ that is significantly larger than the results discussed below.)
We find that the lowest $\chi^2$ for $\alpha/\beta \in [0.1,0.85]$ always
corresponds to $| {\rm cos} \theta |$ near one; this result is illustrated in both
Fig.~\ref{fig:dipolefig} and Table~\ref{dipoletable}.   For any fixed choice of
$\cos\theta$ such that $|\cos\theta|\agt 0.85$, one can find a value of $\alpha/\beta$ that
provides a good fit to the data, {\em i.e.,} the qualitative behavior of $\chi^2$ as a
function of $\alpha/\beta$ is the same as the $\cos\theta=1$ curve in Fig.~\ref{fig:dipolefig}.
\begin{table}[b]
\begin{center}
\begin{tabular}{lcccccccc}
\hline\hline
$\alpha/\beta$ & $0.20$ & $0.30$ & $0.40$ & $0.50$ & $0.60$ & $0.70$ &
$0.80$ & $0.90$ \\ \hline
$\cos\theta$   & $0.89$ & $0.93$ & $0.97$ & $0.99$ & $1.00$ & $-1.00$&
$-0.99$& $-0.88$ \\
$\chi^2$       & $1.33$ & $1.32$ & $1.34$ & $1.35$ & $1.40$ & $1.46$ &
$1.41$ & $20.7$  \\
\hline\hline
\end{tabular}
\caption{Minimum of $\chi^2$ in $\cos\theta$ for fixed $\alpha/\beta$,
assuming dipole wave functions.\label{dipoletable}}
\label{tabledipole}
\end{center}
\end{table}
Table~\ref{dipoletable} illustrates that the lowest $\chi^2$ over a wide range of
$\alpha/\beta$ (corresponding to some $\cos\theta$ value near one) varies only weakly in
$\alpha/\beta$. The best fit point globally is at $\alpha/\beta=0.266$ and $\cos\theta=0.917$.  In
particular, for $\alpha=0.244$~fm and $\beta=0.917$~fm, one finds
\begin{eqnarray}
r_{Ep} = 0.895\mbox{ fm}\ ,  \,\,\,\,\,&&\,\,\,\,\,  r_{Mp}=0.896\mbox{ fm}\ , \nonumber \\
r_{Mn} = 0.862\mbox{ fm}\ , \,\,\,\,\,&&\,\,\,\,\,   r_{En}^2= -0.119\mbox{ fm}^2 \,\,\,.
\end{eqnarray}

\subsection{Gaussian Model}

%%%%%%%%%%%%%%%%%%%%%%%%%%%%
\begin{figure}[t]
\begin{center}
\epsfxsize 3.5 in \epsfbox{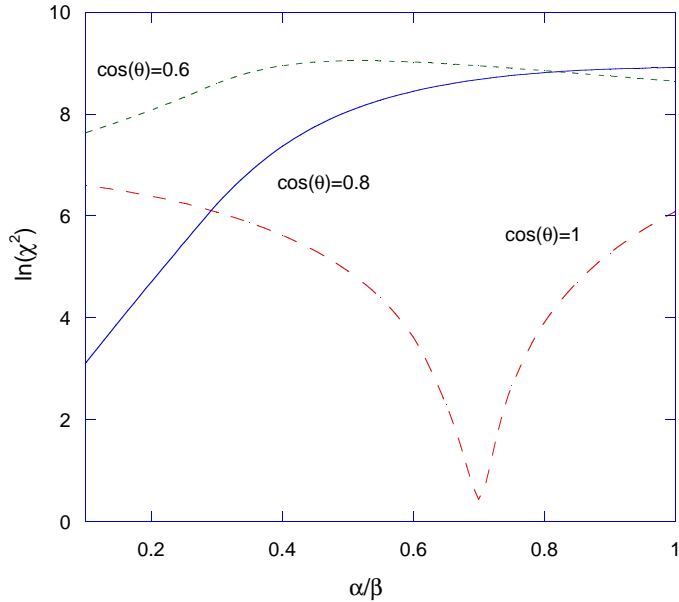} \caption{Plot of $\ln(\chi^2)$ for fixed mixing angle,
assuming Gaussian wave functions.\label{fig:gausscase}}
\end{center}
\end{figure}
%%%%%%%%%%%%%%%%%%%%%%%%%%%%
The form of the momentum-space wave functions that we have just considered was motivated by the asymptotic
behavior of the form factors at large momentum transfer.   However, the charge radii are evaluated at
zero momentum transfer and one might anticipate that other choices for the momentum-space wave functions
(with differing high-$Q^2$ behavior) should produce qualitatively similar results.  We demonstrate
this point by considering Gaussian wave functions. Although Gaussian wave functions lead naively to form
factors that fall off too quickly at high $Q^2$, this limit is also outside the range of validity of our
nonrelativistic treatment; we do not let this point concern us further.  We simply use the
Gaussian case to illustrate the model-independence of our charge radii results.  To proceed, we
take
\begin{equation}
\phi(abc) = \left(\frac{2\alpha\beta}{\pi}\right)^{3/2}
\exp[-\alpha^2 ({p}_{-}^{(abc)})^2] \exp[-\beta^2 ({p}_{+}^{(abc)})^2]  \,\,\,.
\end{equation}
Evaluation of the electric and magnetic charge radii proceeds exactly as described in
Section~\ref{formal}; the only notable difference is that the necessary integration can be done
more easily.  In particular, the master integral of Eq.~(\ref{eq:masteri}) evaluates to
\begin{eqnarray}
{\cal I} ( \gamma , \delta ) & = & \left[
\frac{4 \alpha^2 \beta^2}{ ( \alpha^2 + \beta^2 )^2 -
( \alpha^2 - \beta^2 )^2 \cos^2 \gamma } \right]^{3/2} \nonumber \\
& \times & \exp \left\{ -\frac{Q^2}{3} \frac {\alpha^2 \beta^2
\left[ 2 (\alpha^2 \! + \beta^2) + ( \alpha^2 \! - \beta^2)
\left( \sin 2\gamma \sin 2\delta \! - 2 \! \cos^2 \! \gamma \cos 2\delta
\right) \right] }{ ( \alpha^2 + \beta^2 )^2 - ( \alpha^2 - \beta^2 )^2
\cos^2 \gamma } \right\} .
\end{eqnarray}
The qualitative features of the results are the same as in the previous case:
Good fits are only obtained for $\cos\theta$ near $\pm 1$.  For example, for $0.2 \leq \alpha/\beta
\leq 1$, we find that the best fit values all occur for $\cos\theta>0.85$.  This feature of our solutions
is evident in Fig.~\ref{fig:gausscase}, which shows $\ln(\chi^2)$ (using the data selection in
Ref.~\cite{HdJ}) as a function of $\alpha/\beta$, for three choices of $\cos\theta$.  Only for
$\cos\theta$ near one does the curve develop a deep local minimum that corresponds to a good fit
to the data.  For any $\cos\theta>0.85$, there is an $\alpha/\beta$ that provides a low $\chi^2$
fit to the data.  The set of these $\alpha/\beta$ values spans a wide range, as shown
in Table~\ref{tablegauss} below.
\begin{table}[ht]
\begin{center}
\begin{tabular}{lcccccccc}
\hline\hline
$\alpha/\beta$          & $0.20$ & $0.30$ & $0.40$ & $0.50$ & $0.60$ & $0.70$ & $0.80$ & $0.90$ \\ \hline
$\cos\theta$ & $0.85$ & $0.90$ & $0.95$ & $0.98$ & $1.00$ & $1.00$ & $-1.00$ & $-0.99$ \\
$\chi^2$       & $1.38$ & $1.34$ & $1.34$ & $1.36$ & $1.40$ & $1.54$ & $1.55$ & $1.65$  \\
\hline\hline
\end{tabular}
\caption{Minimum of $\chi^2$ in $\cos\theta$ for fixed $\alpha/\beta$, assuming Gaussian wave functions.}
\label{tablegauss}
\end{center}
\end{table}
As in our previous model, these best $\chi^2$ values remain relatively flat in $\alpha/\beta$ until one
gets very close to $\alpha/\beta=1$ from below.  Thus, our qualitative conclusion remains the same:  the
current data does not strongly favor either the no-diquark or small-diquark solutions.
The best fit point globally is in this case at $\alpha/\beta=0.360$ and $\cos\theta=0.927$.  For
$\alpha=0.311$~fm and $\beta=0.865$~fm, one finds
\begin{eqnarray}
r_{Ep} = 0.895\mbox{ fm}\ ,  \,\,\,\,\,&&\,\,\,\,\,  r_{Mp}=0.897\mbox{ fm}\ , \nonumber \\
r_{Mn} = 0.863\mbox{ fm}\ , \,\,\,\,\,&&\,\,\,\,\, r_{En}^2= -0.119\mbox{ fm}^2 \,\,\,,
\end{eqnarray}
where the overall scale of the parameters has been chosen so that $r_E^p$ reproduces
the correct experimental central value.

In both the dipole and Gaussian models we find that the most constraining feature of the experimental
data is the smallness of the neutron electric charge radius relative to the other radii. The increase
in $\chi^2$ away from the best-fit solutions comes almost entirely from difficulty in accommodating
$r_{En}^2$. The value of this datum drives the fits to large values of $\cos\theta$, and toward particular values
of $\alpha/\beta$, where the model can best accommodate the two distinct length scales represented by the
data. We find that our results do not change qualitatively if we change the experimental value of
$r_{En}^2$ to another small number of either sign, or even set it identically to zero.

It is also worthwhile to note that more distinct global minima can be obtained if
one reduces the uncertainties on the charge radii.  For example, if one arbitrarily shrinks
the uncertainty of the ratio $r_{Mp}^2/r_{Ep}^2$ by a factor of two in our $\chi^2$ function, one finds
that a new global minimum appears near $\alpha/\beta \approx 0.7$, and that the depth of this minimum
is enhanced in comparison to the results presented in Table~\ref{tablegauss}. (For example, the
difference between the $\chi^2$ at the minumum and the $\chi^2$ at $\alpha/\beta \approx 0.2$ is
increased by an order of magnitude when the uncertainties are reduced in this way.) The significant spread
in central values of the various existing estimates of the charge and magnetic
radii~\cite{HdJ,sick,Kelly,codata} suggests that an improvement in our input data is possible. In
this case, a more definitive result on diquark correlations could plausibly be obtained from the
charge radii alone, without consideration of the higher momentum behavior of the form factors.

\section{Conclusions} \label{concl}

The possibility of significant spatial correlations between quark pairs within three-quark baryons
has been met with renewed interest in recent years.  In this Letter, we have considered whether precise
measurements of the electric and magnetic charge radii are sufficient to diagnose whether any
correlation identifiable as a diquark is present within the nucleon.  Since charge radii are
evaluated at zero momentum transfer, we have considered nonrelativistic quark models that
properly encode the total antisymmetry of the spin-flavor-momentum wave functions.  Unlike some
of the other diquark models in the literature, ours allows us to continuously interpolate between
a small diquark (where such antisymmetrization is not necessary) and a mild spatial correlation
that is not significantly smaller than the size of the nucleon.   Working in momentum space, we
have considered wave functions of dipole and Gaussian form.  While the former has a more realistic
high-$Q^2$ fall-off, our results for the charge radii in each case agree qualitatively, suggesting that
our conclusions are relatively model independent.  We find a good fit to the data for a range of spatial
correlations, provided that the correlated quarks are primarily in a scalar diquark state, as we have
defined it. Local minima of comparable $\chi^2$ are present for both small-diquark and no-diquark
solutions, suggesting that a conservative assessment of the experimental uncertainty of the nucleon charge
radii does not yet allow a definitive statement on the existence of a diquark in the nucleon.  However,
the spread in existing determinations of the charge radii suggests that there is room for a reduction
in the uncertainty of our input parameters; in this case, it should be possible, in principle, to determine
whether or not small diquark solutions are favored, without reference to the higher-$Q^2$ behavior of
the form factors.  In the meantime, the overall results of our analysis suggest that a fully relativistic
extension of our approach ({\em i.e.}, evaluation of the nucleon form factors assuming
properly symmetrized wave functions that can smoothly interpolate between spatially correlated and
uncorrelated quark pairs) is well motivated and worthy of further investigation.

{\it Acknowledgments.}  We thank the N.S.F.\ for support through grant Nos.\ PHY-0245056 (C.E.C.),
PHY-0456525 (C.D.C.\ and H.J.K.), and PHY-0456520 (R.F.L.).

\end{document}